\documentclass[fleqn,usenatbib]{mnras}

\usepackage{newtxtext,newtxmath}

\usepackage[T1]{fontenc}
\usepackage{adjustbox}
\usepackage{tablefootnote}

\DeclareRobustCommand{\VAN}[3]{#2}
\let\VANthebibliography\thebibliography
\def\thebibliography{\DeclareRobustCommand{\VAN}[3]{##3}\VANthebibliography}

\usepackage{graphicx}	
\usepackage{adjustbox}
\usepackage{arydshln}
\usepackage{longtable}
\usepackage{lscape}

\title[Extreme N-emitters in DESI DR1 SFGs at z$<$0.5]{The origin of extreme N-emitters in star-forming galaxies at z$<$0.5 with DESI DR1}

\author[Souradeep Bhattacharya]{
Souradeep Bhattacharya,$^{1}$\thanks{E-mail: s.bhattacharya3@herts.ac.uk} and
Chiaki Kobayashi$^{1}$
\\
$^{1}$Centre for Astrophysics Research, Department of Physics, Astronomy and Mathematics, University of Hertfordshire, Hatfield AL10 9AB, UK
}

\date{Accepted XXX. Received YYY; in original form ZZZ}

\pubyear{2025}

\begin{document}
\label{firstpage}
\pagerange{\pageref{firstpage}--\pageref{lastpage}}
\maketitle

\begin{abstract}
Extreme nitrogen enhancement relative to oxygen, recently found in very high-redshift galaxies, has been seen in local star-forming galaxies 
displaying high log(N/O) values ($\geq\!-1.1$) at relatively low O abundances, 12+log(O/H)$\leq$8. 
Understanding the physical origins of these extreme N-emitters at low redshifts enables us to better constrain chemical enrichment mechanisms that drove such high log(N/O) values in the early Universe. 
With direct N and O abundances derived for 944 SFGs with spectroscopic observational data from the Dark Energy Spectroscopic Instrument Data Release 1 (DESI DR1), we report the discovery of 19 extreme N-emitters at low-z (z$<$0.5). 
Our sample of N-emitters represents a five-fold increase in their known number at low-z with 12+log(O/H)$\leq$8, and statistically, $2.21\pm0.91$\% of DESI DR1 SFGs with reliable O and N abundances obtained directly, are extreme N-emitters.
The sample spans a mass range of $\sim 10^7$--$10^{10}$~M$_{\odot}$ with 12+log(O/H) range of $\sim$7.1--8.2, and the N-emitter fraction is found to increase with increasing stellar mass and decreasing metallicity. The most extreme N-emitter in our sample has log(N/O)=$-0.53\pm0.13$, while also having the lowest 12+log(O/H)=$7.08\pm0.09$ and the highest stellar mass, log(M$_{*}$/M$_{\odot}$)=9.95 ± 0.13 among our sample. With galactic chemical evolution models, we show that sustained N-enhancement by asymptotic giant branch stars, in conjunction with presence of outflows during the evolution of the galaxy, can well explain the high log(N/O) of low-z extreme N-emitters. While single starbursts with outflow are sufficient to explain lower-mass N-emitters, more massive ones require a dual starburst scenario where a secondary starburst is triggered by inflow of gas.
\end{abstract}





\section{Introduction} 
\label{sec:intro}

JWST/NIRSpec \citep[][]{Jakobsen22} observations of high-redshift (z$>$1) star-forming galaxies (SFGs) have enabled direct determination (via reliable detections of temperature sensitive auroral lines) of their inter-stellar medium (ISM) abundances of various light elements \citep[e.g.][]{Isobe23, Rogers24, Arellano-Cordova25, Bhattacharya25, Stiavelli25, Cataldi25}. Of particular interest are those high-z SFGs that have been found to display high log(N/O) values (termed extreme N-emitters or N-loud galaxies) at relatively low O abundances, 12+log(O/H)$\leq$8 \citep[][]{Bunker23,Cameron23,Isobe23,Sanders24, Marques-Chaves24, Schaerer24, Castellano24, Curti24, Arellano-Cordova25, Welch24, Welch25, Topping24,Topping25,Stiavelli25, Zhang25}. We adopt a standard definition of extreme N-emitters (hereafter `N-emitters' for brevity) in this work as such SFGs with log(N/O)~$\geq-1.1$. 

Various theoretical scenarios have been suggested to explain the high N/O at low O abundances in these high-z SFGs. These include dilution by pristine gas between intermittent starbursts \citep[][]{Kobayashi24}, enrichment from winds of Wolf-Rayet (WR) stars possibly coupled with clustered star-formation \citep{Isobe23,Marques-Chaves24, Fukushima24, Rivera-Thorsen24},  pollution from Population III star-formation \citep[][]{Rossi24, Nandal24, Tsiatsiou24,Senchyna24}, tidal disruption of stars from encounters with black holes \citep[][]{Zhang25}, ejecta from very massive stars (VMS, 100--1000 M$_{\odot}$) formed through collisions in dense clusters \citep[][]{Vink23}, and supermassive stars (SMS, $\sim10^4$ M$_{\odot}$; \citealt{Charbonnel23, Nagele23,Nandal24b}). We note that some identified N-emitters at high-z are shown to host an active galactic nucleus (AGN) (e.g. GS~3073, \citealt[][]{Ji24}). However, in this work we are only interested in N-emitters that are SFGs.

At low-redshifts, N/O has been routinely determined for individual bright HII regions within the Milky Way (MW; e.g. \citealt{Rubin69, Peimbert69, Esteban18}) and nearby galaxies (e.g. NGC 6822--\citealt{Peimbert70}; M31,M101: \citealt{Esteban20}),  for the integrated spectra of galaxies \citep[e.g.][]{Izotov06}, for damped Lyman-$\alpha$ (DLA) systems \citep[e.g.][]{Battisti12}, as well as for stars in the MW \citep[e.g.][]{Ecuvillon04}. These determinations have shown that N/O is nearly constant with increasing O abundance when 12+log(O/H)$\approx$7.5--8, and increase thereafter with increasing O abundance \citep[][]{Dopita16,Nicholls17}. The increase is associated with enrichment from asymptotic giant branch (AGB) stars from initially $\sim$4--$7M_\odot$ stars \citep[][]{Vincenzo18b}. At 12+log(O/H)$\leq$7.5, N/O determinations remain sparser, although many blue compact dwarfs 
show slightly enhanced N relative to the near-constant values determined when 12+log(O/H)$\approx$7.5--8 \citep[e.g.][]{Zinchenko24}.


\begin{figure*}
\centering
\includegraphics[width=\textwidth]{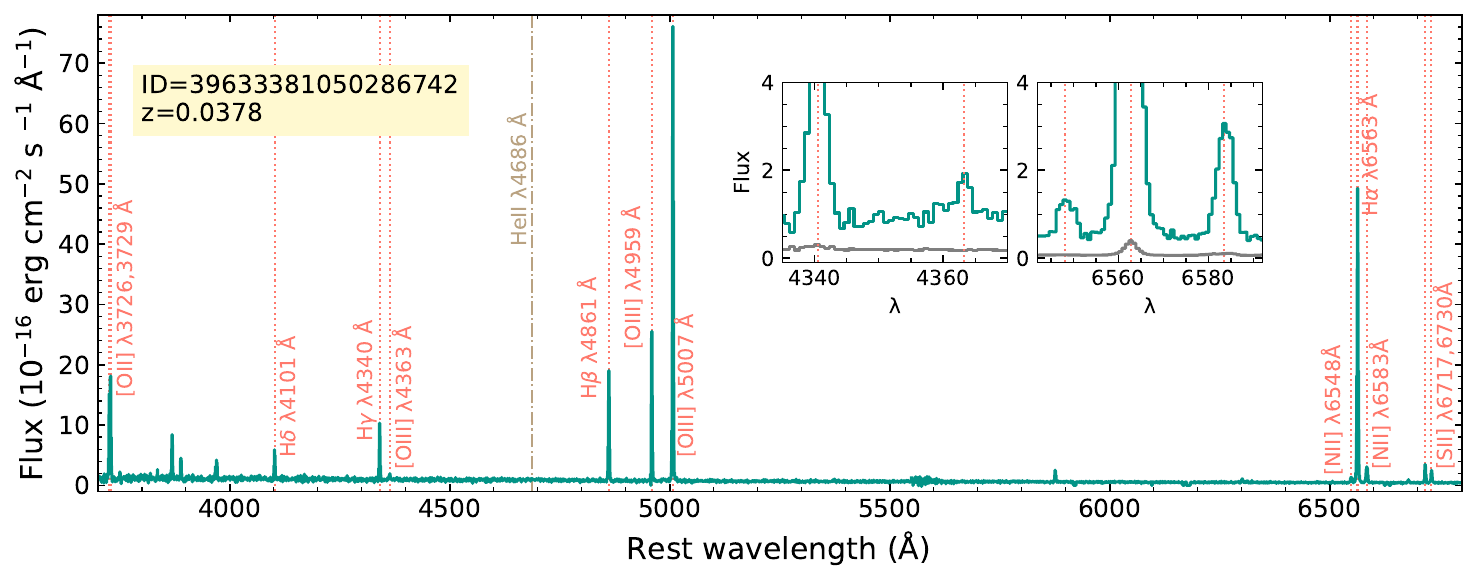}
\caption{An example of the flux and wavelength calibrated spectrum (green) of a SFG observed by DESI (for the N-emitter No. 1 in Table~\ref{tab:attributes}; its DESI ID and redshift are noted). The emission lines utilised in this work are labelled in red. The expected position of the HeII$\lambda$ 4686 \AA~line is marked (brown) but is not observed. The insets show the zoomed-in views around the H$\gamma$ and [OIII]$\lambda$ 4363 \AA; and H$\alpha$ and [NII]$\lambda$ 6548, 6583 \AA~lines, respectively, both of which clearly show their detections. The uncertainty in the flux is marked in grey in these insets.} 
\label{Fig:exp_spec}
\end{figure*}


In the Local Volume ($<10$~Mpc), extreme N-emitters are rare with only the nuclear star-forming region in the blue compact dwarf galaxy Mrk~996 showing extremely high log(N/O)\,$\sim\!-0.15$ at low O abundances, 12+log(O/H)\,$\sim$\,7.9 \citep{James09,Telles14}. Homogeneous spectroscopic survey of SFGs with the Sloan Digital Sky Survey (SDSS; \citealt{Aihara11}) allowed direct N \& O abundance measurements for 231 low-z galaxies (out to z$\sim$0.4; \citealt{Izotov06}). However, only a handful of SFGs with log(N/O)~$\geq\!-1.1$ were observed with most of these having 12+log(O/H)\,$\geq\,$8. 

The Dark Energy Spectroscopic Instrument (DESI; \citealt{DESI22}) on the Mayall 4m telescope at Kitt Peak National Observatory has carried out a $\sim$9000 sq. deg. optical survey of galaxies over its first 13 months of operation, with data made publicly available through the Data Release 1 \citep[DESI DR1;][]{DESI25}. This allows a renewed opportunity to conduct a homogeneous spectroscopic search for SFGs to directly determine their N \& O abundances, but now over wider and potentially deeper scales than SDSS. 

In this paper, we obtain direct N \& O abundances for a sample of DESI DR1 SFGs at $z\!<\!0.5$, to identify a homogeneous sample of N-emitters in the local Universe. Using galactic chemical evolution (GCE) models, we then constrain the physical processes that caused the observed high N/O values for the SFGs in our sample, to subsequently gain insight on the origin of N-emitters at high-z. Our sample selection (including AGN exclusion) and direct abundance determination is described in Section~\ref{sec:sample}. We explore the 12+log(O/H) vs log(N/O) plane for DESI DR1 SFGs and identify N-emitters in Section~\ref{sec:analysis}. We 
compare our identified N-emitters with literature N-enriched SFGs, both at low-z and high-z in Section~\ref{sec:disc}. Using GCE models, we draw inferences on the origin of N-emitters Section~\ref{sec:models}. We conclude in Section~\ref{sec:conc}.


\section{Sample Selection and Abundance determination} 
\label{sec:sample}


\subsection{DESI DR1 Value-added Catalogue Data} 
\label{sec:data}

DESI enables multi-object spectroscopy with $5000$ fibers (each 1$''$.5 in diameter) over a 3$^{\circ}$.2 diameter field of view. The light from each fiber is split into the blue (3600--5550\AA; R=2000--3200), red (5550--6560\AA; R=3200--4100) and 
infrared (6560--9800\AA; R=4100--5000) arms of a spectrograph, providing continuous wavelength coverage over the specified optical range. An example DESI spectrum of a SFG is shown in Figure~\ref{Fig:exp_spec}. 

The DESI main survey is intended to be 5-year observational program targeting 30 million pre-selected galaxies across one-third of the night sky in-order to measure redshifts. DESI DR1 consists of all data acquired during the first 13 months of the DESI main survey and uniform reprocessed DESI Survey Validation data that had been previously made publicly available as the DESI Early Data Release \citep[EDR;][]{DESI24}. The DESI DR1 main survey includes high-confidence redshifts for 18.7 million sources, of which 13.1 million are spectroscopically classified as galaxies. 

In this work, we utilize the stellar mass and emission-line value added catalog (VAC\footnote{see \url{https://data.desi.lbl.gov/doc/releases/dr1/vac/stellar-mass-emline} for details.}) from DESI DR1 (Zou et al. 2025 in prep; consistent with \citealt{Zou24} for EDR). This VAC provides the position (RA, DEC), spectroscopic redshift value (z), stellar mass (M$_{*}$), star-formation rate (SFR) and emission-line flux measurements for all sources classified as galaxy in DESI DR1 with reliable redshift measurements (see Table~\ref{tab:sample}).  For each source, following absorption correction through continuum fitting performed by the stellar spectral synthesis code STARLIGHT \citep[][]{Cid05}, a pre-defined set of optical emission lines are measured by a single Gaussian fit. M$_{*}$ is derived using the stellar population modelling software CIGALE \citep[][]{Boquien19} from the broad-band g, r, z, W1, and W2 photometry from the DESI Legacy Imaging Surveys \citep{Dey19} and spectrophotometry of 10 artificial bands generated through convolution with DESI spectra (see \citealt{Zou24} for details). SFR is derived from the flux of the H$\alpha$-line where available. Thus while the tabulated SFR value only refers to the SFR within the region of the galaxy spanned on-sky by the fiber, the tabulated M$_{*}$ more broadly refers to the stellar mass of the source, even if its entire extent is not encompassed by the fiber. 

We only utilize those tabulated emission-line  fluxes for each galaxy from this VAC that are useful for direct O and N abundance measurements, i.e., [OII]$\lambda\lambda$ 3726,3729 \AA, H$\delta$, H$\gamma$, [OIII]$\lambda$ 4363 \AA, H$\beta$, [OIII]$\lambda\lambda$ 4959,5007 \AA, H$\alpha$, [NII]$\lambda$ 6548,6583 \AA~\& [SII]$\lambda\lambda$ 6717,6731 \AA~(see also Figure~\ref{Fig:exp_spec}). These line fluxes and their errors are utilized in selecting our sample of this work in Section~\ref{sec:final}. The abundance determination is discussed in Section~\ref{sec:abund}. 
\
\begin{table}
\caption{Sample selection of DESI DR1 galaxies in this work, with each subsequent sample being a subset of the former.}
\centering
\adjustbox{max width=\columnwidth}{
\begin{tabular}{l|r}
\hline
Sample & No. of galaxies \\
\hline 
All targets in DESI DR1 value-added catalogue &  14,706,085 \\
\hdashline
Those of the above with z$\leq0.4915$; \\
Flux/$\delta\rm_{Flux}$ ([OIII] 4363 \AA, [OIII] 5007 \AA, H$\beta$) $\geq3$;\\
Flux ([NII] 6583 \AA) $>$0 &  9,209\\
\hdashline
Those of the above that are SFGs, following AGN exclusion, \\
with z$\geq0.032$ and O abundances computed (for Figure~\ref{Fig:lit}) & 1,815\\
\hdashline
Those of the above with Flux/$\delta\rm_{Flux}$ ([NII] 6583 \AA) $\geq3$ \\
and N \& O abundances computed (in Figure~\ref{Fig:sample}) & 944 \\
\hdashline
Those of the above with log(N/O)$\geq-1.1$ & 19 \\
\hline
\end{tabular}
\label{tab:sample}
}
\end{table}

\subsection{Sample selection} 
\label{sec:final}

We start from the $\sim$14.7 million targets in DESI DR1 classified as galaxies in the VAC (1st row of Table~\ref{tab:sample}) and make the following selections:
\begin{enumerate}
    \item As the red-most emission-line of interest required for O and N abundance determination ([NII]$\lambda$ 6583 \AA) was only tabulated for galaxies out to z~$=0.4915$, our sample is limited to this redshift. We further restrict our sample to those sources that have Flux/$\delta\rm_{Flux}$ ([OIII] 4363 \AA) $\geq3$ as reliable line flux measurement of the temperature sensitive [OIII]$\lambda$ 4363 \AA~line is required for direct abundance determination. We find that this does not automatically imply that brighter [OIII] 5007 \AA\& H$\beta$ lines have S/N>3. Thus, we additionally impose the selection criteria that Flux/$\delta\rm_{Flux}$ ([OIII] 5007 \AA, H$\beta$) $\geq3$. Given our interest in obtaining N abundances, we impose the selection criteria of Flux ([NII] 6583 \AA)$>$0 (which also automatically selects only those objects with Flux (H$\alpha$)$>$0)\footnote{We find that a number of sources in the VAC have bright H$\alpha$ \& [NII]$\lambda$ 6583 \AA~lines (also verified visually from the released spectra) but have no tabulated errors. Not to exclude such sources, at least in the first instance, we refrain from S/N selection on these lines.}. In this first instance, we thus have 9209 galaxies after our selection (2nd row of Table~\ref{tab:sample}).
    \item Visually checking these galaxies in the DESI Legacy Imaging Survey images\footnote{\url{https://www.legacysurvey.org}}, we find that many lowest redshift sources classified as galaxies, are actually individual HII regions in nearby spiral galaxies, having emission-line spectra. We thus impose an additional selection criteria of z$\geq0.032$ to exclude such HII regions\footnote{A small number of dwarf galaxies could also be unfortunately excluded with this criteria.}. For the remaining sources which are galaxies, we exclude candidate AGN by cross-matching with the DESI DR1  AGN/QSO VAC\footnote{See \url{https://data.desi.lbl.gov/doc/releases/dr1/vac/agnqso} for details.} (Juneau et al. in prep.). AGN candidates have been identified from the full DESI DR1 sample in this AGN/QSO VAC from various emission-line \citep{Kewley01,Kauffmann03,Cid11,Shirazi12,Juneau14,Kai18,Law21} and infrared diagnostics \citep{Jarrett11,Mateos12,Stern12,Assef18,Yao20,Hviding22}. After removing these candidate AGN, the remaining emission-line galaxies in our sample are all SFGs based on the aforementioned emission-line and infrared diagnostics. For these SFGs, we proceed to abundance determination for O \& N abundances (Section~\ref{sec:abund}). Only those galaxies with reliable O abundances are included in the selection. This results in 1815 SFGs (3rd row of Table~\ref{tab:sample}). This sample is used to check the consistency of our determined O abundance with previous works in Section~\ref{sec:lit}.
    \item N abundances are determined for a sub-sample of these SFGs where Flux/$\delta\rm_{Flux}$ ([NII] 6583 \AA) $\geq3$, resulting in 944 SFGs that are studied in this work (4th row of Table~\ref{tab:sample}).
\end{enumerate} 


\subsection{Abundance determination} 
\label{sec:abund}

Our abundance determination procedure using NEAT \citep[Nebular Empirical Analysis Tool;][]{Wesson12} is laid out in detail in earlier papers \citep[][]{Bhattacharya22,Bhattacharya25, Bhattacharya25b}. We briefly describe the procedure here. For each SFG, NEAT utilizes an iterative procedure for calculating the intrinsic balmer decrement, c(H$\beta$), from flux-weighted ratios of H$\alpha$/H$\beta$, H$\gamma$/H$\beta$ and H$\delta$/H$\beta$ (as available), nebular temperature (T$\rm_e$) from the temperature-sensitive [OIII]$\lambda$ 4363 \AA~ line, and electron density (n$\rm_e$) from the density-sensitive [OII]$\lambda\lambda$ 3726,3729 \AA\, and [SII]$\lambda\lambda$ 6717,6731 \AA\, doublets (whichever are observed). We assume the extinction law of \citet{Cardelli89}. For those galaxies where we do not observe the required doublets to determine n$\rm_e$, we assume n$\rm_e$=1000 cm$^{-3}$, which is expected to have negligible impact on the determined abundances \citep[e.g.,][]{Ferland13}. Only those SFGs with c(H$\beta$)$>0$, T$\rm_e < 35000$~K and n$\rm_e < 10000$~cm$^{-3}$ are considered to be reliable and their direct abundances are determined. \citep[see also][]{Bhattacharya22,Bhattacharya25b,Bhattacharya25}.

Direct O and N ionic abundances are determined from the measured fluxes of the O ([OII]$\lambda\lambda$ 3726,3729 \AA,  [OIII]$\lambda\lambda$ 4363,4959,5007~\AA) and N ([NII]$\lambda$ 6548,6583 \AA) lines respectively. Ionisation correction factors (ICF) for O is negligible when lines pertaining to both O$^{2+}$ (i.e, [OIII] $\lambda\lambda$ 5007, 4959, 4363 \AA) and  O$^{+}$ ([OII] $\lambda\lambda$ 3727, 3729 \AA) are observed. As the objective of this work is to check for N-enrichment, we can bypass the need for ICFs by directly obtaining log(N/O) as log(N$^{+}$/O$^{+}$), which is considered to be a reasonable approximation for metal-poor galaxies \citep[][]{Izotov04}.


\begin{figure}
\centering
\includegraphics[width=\columnwidth]{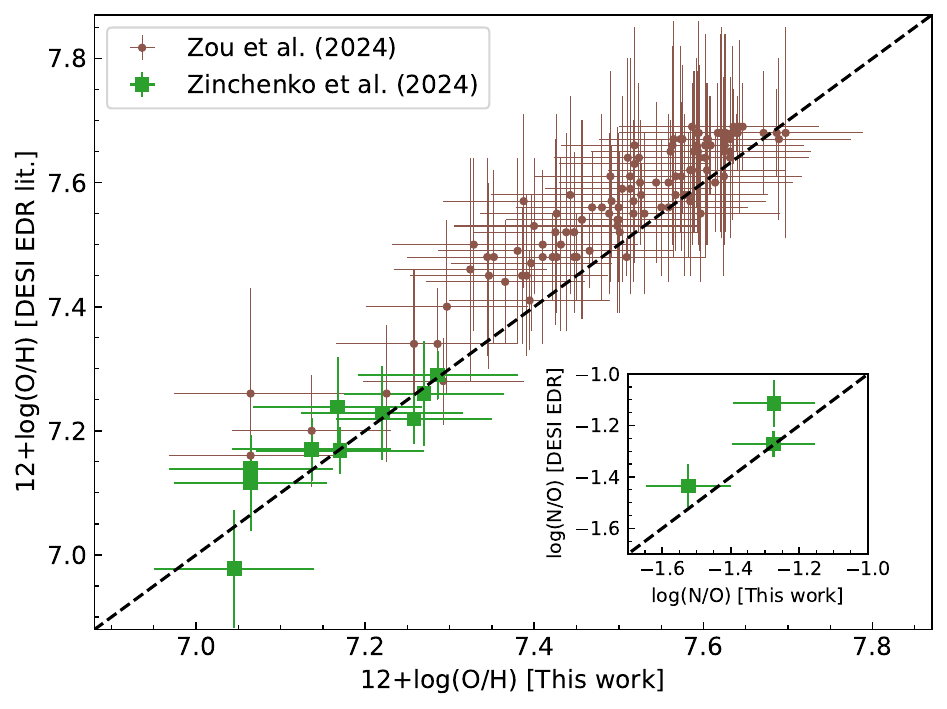}
\caption{12+log(O/H) abundances determined in this work compared to previous determinations (green squares, \citealt{Zinchenko24}; magenta dots, \citealt{Zou24}) from DESI EDR spectra. The 1:1 line is also marked. The inset shows the same for log(N/O).} 
\label{Fig:lit}
\end{figure}

\subsection{Consistency with abundance determinations from DESI EDR} 
\label{sec:lit}

\citet{Zinchenko24} had utilized the DESI EDR spectra to directly determine O abundances for 21 extremely metal-poor galaxies (12+log(O/H)<7.3), 8 of which also had N abundances reported. Only 11 galaxies from their sample (including 3 with N abundances reported) are present in our selection of 1815 galaxies with reliable O abundances. Interestingly, none of these galaxies are among the 944 galaxies in our final selection that has reliable N abundances. This results from S/N>3 considerations in the VAC for the [OIII] 4363 \AA~line, which excludes 10 galaxies from \citet{Zinchenko24} in our sample, and the remaining 11 galaxies are pruned as they all have no errors noted for the [NII] 6583 \AA~line in the VAC. 

Visual confirmation of the DESI DR1 spectra of the galaxies from \citet{Zinchenko24} gave the impression that the required [OIII] 4363 \AA~and [NII] 6583 \AA~lines are present in the spectra as reported by the authors based on the EDR spectra, but these lines have low S/N, with the VAC conservatively assigning slightly higher errors than that determined by the authors. This is consistent with the presence of only 7 of the galaxies from \citet{Zinchenko24} being present in the metal-poor sample of \citet{Zou24}, who had utilized the DESI EDR VAC to report directly determined O abundances for 223 very metal-poor galaxies (12+log(O/H)<7.7) in DESI EDR, 193 of which are within z=0.4915. 

Of these 193 galaxies, only 115 are present in our selection of galaxies with reliable O abundances. The ones excluded from our sample are primarily AGN candidates classified as such with the WISE colour-based AGN candidate selection criteria. While there is considerable overlap between AGN and SFGs in such WISE colour selections \citep[e.g.][]{Jarrett11}, we conservatively exclude all AGN candidates. 

We compare our determined O abundances with those of the 11 galaxies in common with \citet{Zinchenko24} and the 115 in common with \citet{Zou24}. Figure~\ref{Fig:lit} shows that we are consistent with the O abundances determined by \citet{Zinchenko24} within the errors, but have mean O abundances lower by 0.06~dex than those determined by \citet{Zou24}, slightly more than the standard deviation of 0.044~dex, albeit still consistent within error. The offset likely stems from the choice of non-standard, albeit justified, atomic recombination and collision strength data by \citet{Zou24} which is expected to result in their slightly higher O abundances. The abundance determination procedure, albeit using different software, is very similar between this work and that of \citet{Zinchenko24}. It is thus reassuring to determine consistent O abundances, as it bolsters the consistency between the line fluxes measured from the DESI spectra, and those reported in the VAC we use in this work. 

For 3 SFGs, we can further compare the log(N/O) determined in this work with those reported by \citet{Zinchenko24}. We reiterate that these galaxies are not present in our final selection of 944 galaxies with reliable N abundances as their [NII] 6583 \AA~lines flux errors are unreported in the VAC. Nevertheless, we assume a 10\% error on the reported flux in this line (an acceptable value visually checked with the spectra) to enable computation of the N abundances and thereby comparison with the reported values. Inset of Figure~\ref{Fig:lit} shows that our determined log(N/O) values are consistent with those reported by \citet{Zinchenko24} for these 3 SFGs.


\begin{figure}
\centering
\includegraphics[width=\columnwidth]{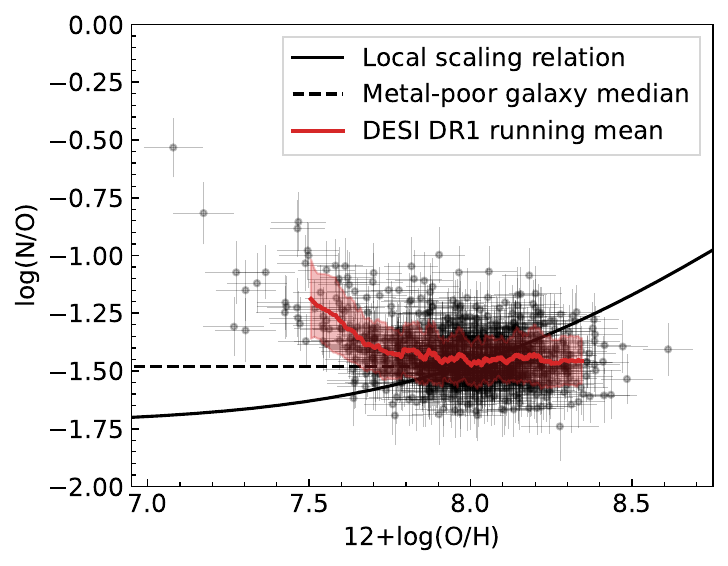}
\caption{12+log(O/H) vs log(N/O) for the 944 DESI DR1 SFGs with our reliable O \& N abundances. The uncertainties are marked with errorbars. 
The running mean in log(N/O) as a function of 12+log(O/H) is shown in red, with the 1~$\sigma$~uncertainty shaded. The local scaling relation \citep[solid line]{Dopita16,Nicholls17} is also marked, along with the median log(N/O) value for metal-poor galaxies computed by \citet[dashed line]{Zinchenko24}.} 
\label{Fig:sample}
\end{figure}

\section{Analysis} 
\label{sec:analysis}


\subsection{The log(N/O) vs 12+log(O/H) plane of SFGs} 
\label{sec:no}

For the 944 SFGs with reliable O and N abundances directly determined in this work, we present their distribution in the 12+log(O/H) vs log(N/O) plane in Figure~\ref{Fig:sample}. The running mean of log(N/O) is computed as a function of increasing 12+log(O/H) in a fixed window of 50 SFGs. 
The N/O increase at 12+log(O/H)$>8$, as per the local scaling relation\footnote{This was computed from MW stars and HII regions in nearby galaxies  \citep[][see also the references for the original observations in Section~\ref{sec:intro}]{Dopita16,Nicholls17}.}, is not clearly seen for our sample. This is probably due to a selection effect of our sample where the [OIII]$\lambda$ 4363 \AA~line flux (for the same [OIII]$\lambda$ 5007 \AA~line flux) is inversely correlated with metallicity. So for any flux-limited spectroscopic survey, the [OIII]$\lambda$ 4363 \AA~line is detected with higher S/N in relatively metal-poor galaxies \citep{Curti20}. However, this is not a problem in this work for finding the low-z counterparts of high-z N-emitters, as most SFGs that are N-emitters at high-z are at 12+log(O/H)$<8$. 

Additionally, \citet{Zinchenko24} made a compilation of metal-poor (12+log(O/H)$<$8) low-z SFGs having direct N \& O abundance determinations from the literature, and computed their mean log(N/O) as $-1.48$, shown with the dashed line in Fig~\ref{Fig:sample} (see also \citealt{Cataldi25} for a very similar relation).  We find that a number of our SFGs, especially those with  12+log(O/H)$<$7.6, have log(N/O)~$>-1.48$.


\begin{figure}
\centering
\includegraphics[width=\columnwidth]{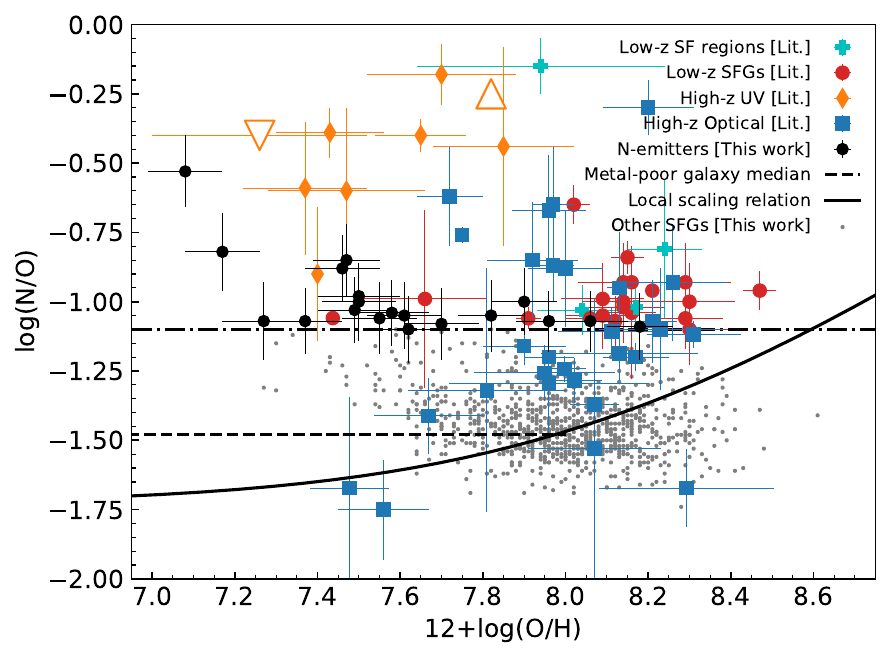}
\caption{12+log(O/H) vs log(N/O) for the DESI DR1 944 SFGs with our reliable O \& N abundances. 19 N-emitters (log(N/O)\,$\geq\!-1.1$; those above dot-dashed line) are marked in black (with their determined uncertainties) while the others are marked in grey. Different groups of literature sources are shown for comparison (see  Section~\ref{sec:disc}). Note that the upward triangle shows the lower limit for GNz11 while the downward triangle shows the upper limit for GHz2. } 
\label{Fig:enrich}
\end{figure}


\subsection{Extreme N-emitting SFGs in DESI DR1} 
\label{sec:enrich}

We now make a selection of extreme N-emitters identified in SFGs at z$<0.5$ from DESI DR1 in this work with the definition of log(N/O)$>-1.1$ (Section~\ref{sec:intro}). The selection criteria is empirically made to allow for comparison with suitably large numbers of literature sources, both high-z and low-z ones, that have previously been studied for N-enhancement (discussed further in Sections~\ref{sec:low-z}~\&~\ref{sec:high-z}). 

\begin{figure*}
\centering
\includegraphics[width=0.97\textwidth]{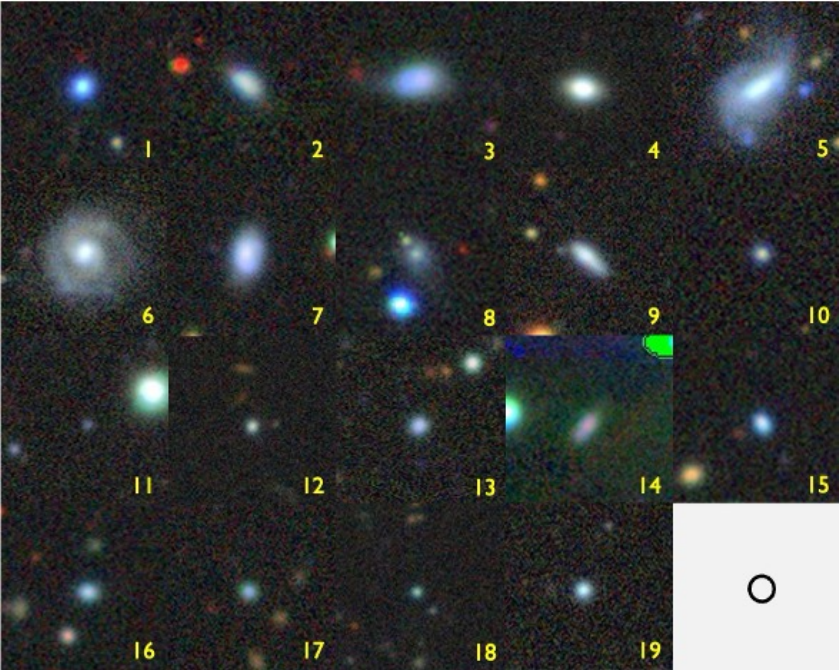}
\caption{RGB image cutouts ($10' \times 10'$) from data release DR10 of the DESI legacy imaging surveys for the 19 SFGs identified in this work having considerably enriched N-abundances (log(N/O)$\geq$-1.1). Their sl. no. from Table~\ref{tab:attributes} is marked. The bottom-right panel shows the DESI fiber on-sky angular extent.} 
\label{Fig:stamps}
\end{figure*}

As a result, we find 19 extreme N-emitters identified in our sample with log(N/O)$>-1.1$ (the last row of Table~\ref{tab:sample}). They are highlighted in the 12+log(O/H) vs log(N/O) plane in Figure~\ref{Fig:enrich} (black dots with errorbars) and their RGB image cutouts are shown in Figure~\ref{Fig:stamps}. While most of these SFGs are completely within the on-sky angular extent of the DESI fiber, some of them at low-redshifts only have their central regions within the fiber. Their IDs, positions, redshifts, stellar masses, SFRs and computed 12+log(O/H) and  log(N/O) are listed in Table~\ref{tab:attributes}. Figure~\ref{Fig:exp_spec} shows the spectrum of the  N-emitter No. 1 in Table~\ref{tab:attributes} with clear detection of the auroral [OIII]$\lambda$ 4363 \AA~line as well as the [NII]$\lambda$ 6548, 6583 \AA~lines. The spectra for the other 18 N-emitters are presented in Figures~\ref{Fig:app_spec_1}~\&~\ref{Fig:app_spec_2} and discussed in Appendix~\ref{appendix: spectra}.

Given the small number of such galaxies identified in our sample of 944 SFGs with reliable O and N abundances, we compute the fraction and uncertainty (95\% confidence interval) in fraction of N-emitters, using the binomial proportion confidence-interval obtained with the Wilson score interval method \citep{Wilson27}. We find that $2.21\pm0.91$\% of DESI DR1 SFGs, with reliable O and N abundances obtained directly, are N-emitters.

It is particularly notable that the most extreme N-emitter in our sample (No. 6 in Table~\ref{tab:attributes} with log(N/O)=$-0.53\pm0.13$) also has the lowest 12+log(O/H) with $7.08\pm0.09$ and the highest stellar mass, log(M$_{*}$/M$_{\odot}$)=9.95 ± 0.13, with the abundance values representative of its central regions (No. 6 in Figure~\ref{Fig:stamps}). The SFGs with the second and third highest log(N/O) values (No. 4 and No. 14 in Table~\ref{tab:attributes} with log(N/O)=$-0.82\pm0.13$~\&~$-0.85\pm0.13$) are also quite massive with log(M$_{*}$/M$_{\odot}$)=9.48 ± 0.11~\&~9.35 ± 0.16 respectively, being also amongst the SFGs with the lowest O abundances in our sample with  12+log(O/H)=$7.17\pm0.09$~\&~$7.47\pm0.08$ respectively. The lowest mass galaxy in our sample (No. 7 in Table~\ref{tab:attributes} with log(M$_{*}$/M$_{\odot}$)=7.03 ± 0.17) also has the fourth highest  log(N/O)=$-0.88\pm0.12$, with a low 12+log(O/H)=$7.46\pm0.09$.


\begin{table*}
\caption{Catalogued attributes and determined abundances of N-enriched galaxies (log(N/O)$\geq$-1.1) identified in DESI DR1.}
\centering
\adjustbox{max width=\textwidth}{
\begin{tabular}{l|l|c|c|c|c|c|c|c}
\hline
No. & DESI TARGET ID & RA & DEC & z & log(M$_{*}$/M$_{\odot}$) & log(SFR) & 12+log(O/H) & log(N/O)\\
\hline 
1 & 39633381050286742 & 191.859176 & 59.565488 & 0.0378 & 7.69 ± 0.18 & -0.37 ± 0.17 & 7.96 ± 0.08 & -1.07 ± 0.11 \\
2 & 39633510662668308 & 100.838738 & 72.678735 & 0.0419 & 8.73 ± 0.14 & -0.42 ± 0.24 & 7.37 ± 0.09 & -1.07 ± 0.12 \\
3 & 39632930795947980 & 199.982441 & 32.591007 & 0.0514 & 9.06 ± 0.14 & -0.2 ± 0.22 & 7.49 ± 0.09 & -1.03 ± 0.12 \\
4 & 39627533284542216 & 58.21239 & -10.411328 & 0.0581 & 9.48 ± 0.11 & -0.15 ± 0.22 & 7.17 ± 0.09 & -0.82 ± 0.13 \\
5 & 39628209783834902 & 157.065201 & 17.867498 & 0.069 & 7.75 ± 0.03 & 0.39 ± 0.03& 7.9 ± 0.08 & -1.0 ± 0.12 \\
6 & 39627850667526584 & 330.368151 & 2.403119 & 0.094 & 9.95 ± 0.13 & 0.93 ± 0.23 & 7.08 ± 0.09 & -0.53 ± 0.13 \\
7 & 39627697579625487 & 204.416316 & -3.658184 & 0.0958 & 7.03 ± 0.17 & -0.99 ± 0.21 & 7.46 ± 0.09 & -0.88 ± 0.12 \\
8 & 39633071447737239 & 118.114262 & 39.782491 & 0.096 & 9.11 ± 0.12 & -0.27 ± 0.25 & 7.58 ± 0.09 & -1.04 ± 0.12 \\
9 & 39627836318813655 & 195.157633 & 2.079406 & 0.0984 & 9.37 ± 0.14 & 0.3 ± 0.26 & 7.82 ± 0.08 & -1.05 ± 0.12 \\
10 & 39633368454794121 & 181.353369 & 58.415714 & 0.1134 & 8.95 ± 0.15 & -0.08 ± 0.28 & 7.5 ± 0.09 & -1.0 ± 0.14 \\
11 & 39627914379004562 & 173.883709 & 5.295752 & 0.1186 & 7.44 ± 0.22 & -0.37 ± 0.25 & 7.5 ± 0.1 & -0.98 ± 0.12 \\
12 & 39627646585280675 & 36.837817 & -5.867579 & 0.1257 & 8.0 ± 0.22 & 0.46 ± 0.16 & 8.18 ± 0.08 & -1.09 ± 0.12 \\
13 & 39628035053326862 & 228.697273 & 10.164711 & 0.127 & 8.76 ± 0.17 & 0.22 ± 0.26 & 7.27 ± 0.1 & -1.07 ± 0.13 \\
14 & 39633203069193270 & 98.286538 & 47.251371 & 0.1336 & 9.35 ± 0.16 & 1.18 ± 0.18 & 7.47 ± 0.08 & -0.85 ± 0.13 \\
15 & 39633039701053482 & 214.447961 & 37.948368 & 0.1359 & 9.21 ± 0.16 & 0.48 ± 0.25 & 7.7 ± 0.09 & -1.08 ± 0.13 \\
16 & 39633058315373292 & 188.542699 & 38.966799 & 0.1659 & 9.27 ± 0.14 & 0.38 ± 0.24& 7.62 ± 0.09 & -1.1 ± 0.12 \\
17 & 39627911006784144 & 332.289797 & 5.081961 & 0.1734 & 9.01 ± 0.17 & 0.23 ± 0.27 & 7.55 ± 0.09 & -1.06 ± 0.12 \\
18 & 39627869588029634 & 18.529157 & 3.598735 & 0.2594 & 7.99 ± 0.14 & 0.54 ± 014 & 8.06 ± 0.09 & -1.07 ± 0.11 \\
19 & 39627763539249582 & 177.116195 & -1.030106
& 0.269 & 9.33 ± 0.14 & 0.8 ± 0.21 & 7.61 ± 0.09 & -1.05 ± 0.12 \\
\hline
\end{tabular}
\label{tab:attributes}
}
\end{table*}

\begin{figure*}
\centering
\includegraphics[width=\textwidth]{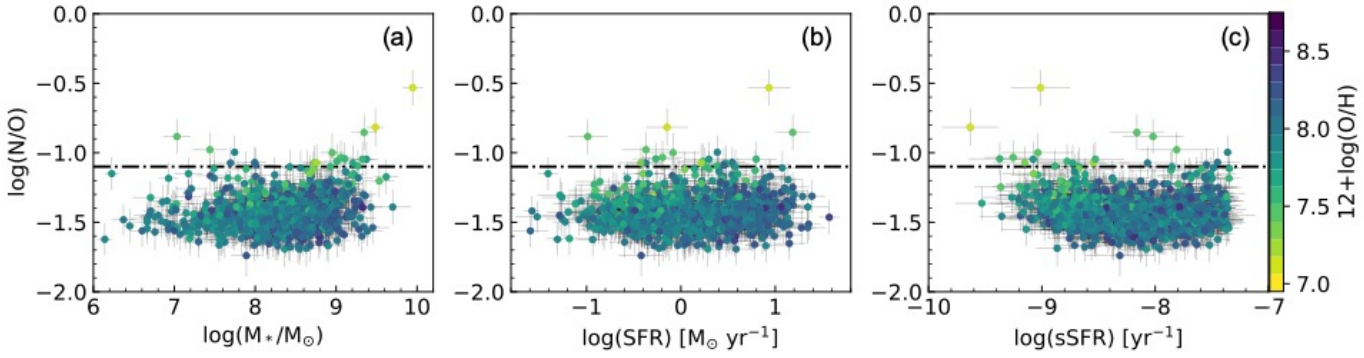}
\caption{(a) Stellar mass vs log(N/O) for the DESI DR1 SFGs with reliable O and N abundances. The uncertainties are marked. The DESI DR1 SFGs are coloured by their 12+log(O/H). The dash-dot line shows log(N/O)=-1.1, the demarcation line of N-emitters. (b) Same as (a), but now showing SFR vs log(N/O). (c) Same as (a), but now showing sSFR vs log(N/O).} 
\label{Fig:mass_sfr}
\end{figure*}


\begin{figure*}
\centering
\includegraphics[width=\textwidth]{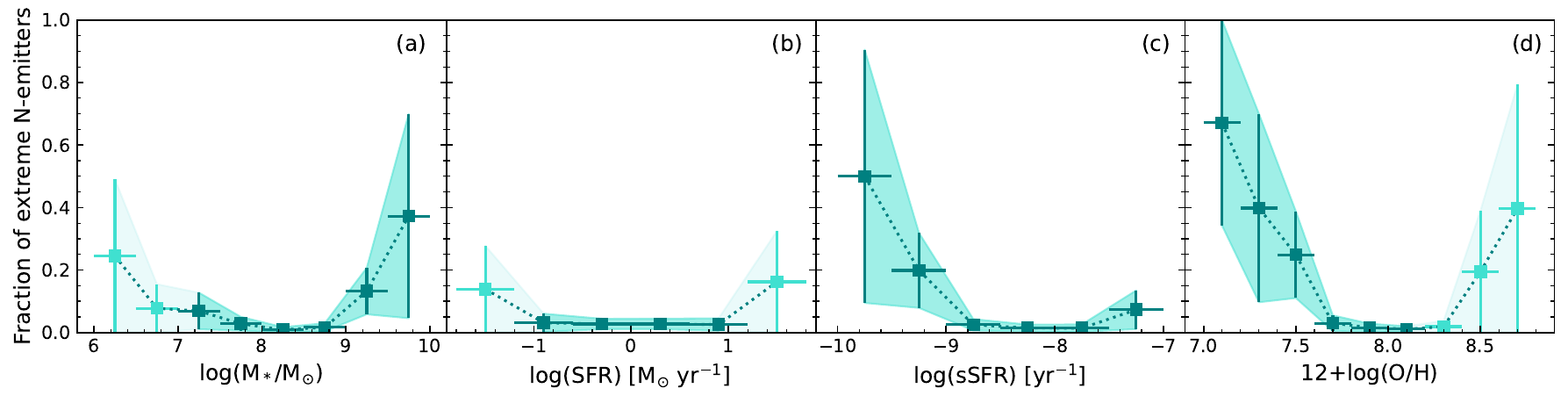}
\caption{The fraction of N-emitters as a function of binned (a) stellar mass, (b) SFR,(c) sSFR \& (d) 12+log(O/H). The fractions and uncertainties (95\% confidence interval) for the fraction of  N-emitters in each bin is computed using the binomial proportion confidence-interval obtained with the Wilson score interval method \citep{Wilson27}. Points marked with darker shade have N-emitters identified in our sample in the respective parameter bin, while those with lighter shade have no N-emitters identified (see Section~\ref{sec:mass} for more details).} 
\label{Fig:frac}
\end{figure*}


\subsection{Relationship with other physical quantities}
\label{sec:mass}

Figure~\ref{Fig:mass_sfr}a shows that our sample of the DESI DR1 SFGs with reliable O \& N abundances consists primarily of relatively lower-mass SFGs with majority having $M_{*}\sim10^8$--$10^9$~M$_{\odot}$. The selection of metal-poor galaxies in our sample (see Section~\ref{sec:no}) in conjunction with the mass--metallicity relation (MZR) of galaxies, whereby galaxies on average show increasing O abundances with increasing mass \citep[e.g.][]{Pagel81, Tremonti04, Curti20}, results in very few massive galaxies being present in our sample. This is still the case although the masses of SFGs in our sample are higher than those in the literature sample compiled by \citet{Zinchenko24}. 

Figure~\ref{Fig:frac}a clearly shows a near constant fraction of N-emitters at $10^7$--$10^9$M$_{\odot}$, with an increase at higher masses.  We find, statistically, that $14.99\pm7.97$\% and $37.24\pm32.69$\% of SFGs are N-emitters with $M*=10^9$--$10^{9.5}$M$_{\odot}$ and $10^{9.5}$--$10^{10}$M$_{\odot}$ respectively. At the lowest masses, given the lack of N-emitters in our sample, we can statistically say that only up to $\sim$15\% and $\sim$50\% of SFGs may be N-emitters with $M_*= 10^{6.5}$--$10^7$M$_{\odot}$ and $10^6$--$10^{6.5}$M$_{\odot}$ respectively. 

Figure~\ref{Fig:mass_sfr}b shows that N-emitters are present over a wide range of SFR values (log(SFR)$\sim-1.2$ to $1.2$~M$_{\odot}$~yr$^{-1}$) with a near constant fraction ($\sim$1--2.5\%; see Figure~\ref{Fig:frac}b). Beyond this SFR range where we find no N-emitters in our sample, yet we can statistically say that up to $\sim$16\% SFGs may be N-emitters.

Figure~\ref{Fig:mass_sfr}c shows that almost the entire sample of DESI DR1 SFGs with reliable O \& N abundances have high sSFR with the bulk of the sample having sSFR$\sim10^{-9}$--$10^{-7.5}$~yr$^{-1}$, which are typical values for starburst galaxies. While the  N-emitters span the full range of sSFR values, it is clear from Figure~\ref{Fig:frac}c that the lowest sSFR bins hold a higher fraction of N-emitters. This is consistent with the increased fraction of N-emitters at higher M$_*$ but spanning a range of SFR values. 

Figure~\ref{Fig:frac}d shows the fraction of N-emitters as a function of binned 12+log(O/H). As already shown in Figure~\ref{Fig:sample}, it is clear that the fraction of  N-emitters increases with decreasing O abundance.


\section{Comparison with literature extreme N-emitting Star-forming galaxies} 
\label{sec:disc}

\subsection{Comparison with low-redshift extreme N-emitting SFGs} 
\label{sec:low-z}

Enhancement of N/O in metal-poor SFGs has been previously reported (e.g. \citealt{Guseva11, Perez-Montero11, Kumari18}). Various mechanisms can be responsible for the enhanced N/O ratios at low metallicity, including localized nitrogen enrichment of the ISM by WR stars \citep[][]{Kumari18}, inflow of metal-poor gas \citep[][]{Koppen05}, varying star formation efficiency \citep{Perez-Montero11} and initial mass function \citep[IMF;][]{Vincenzo16}.  

Figure~\ref{Fig:enrich} also shows the four known star-forming regions in nearby galaxies that show log(N/O)$>-1.1$ (cyan plus symbol). These are the nucleus of Mrk 996 \citep[][]{James09,Telles14}, the center of NGC 5253 \citep[][]{Monreal-Ibero12}, Region 2 in NGC 4670 \citep[][]{Kumari18} and Region 2 in UM 448 \citep[][]{James13}. All four have been suggested to be N-enriched due to WR stars, with Mrk 996 having the highest log(N/O) computed till date from optical N emission lines. Recently, \citet{Abril-Melgarejo24} showed from studying individual HII regions spanning a range of ages in NGC 5253, that young HII regions with ages $\sim$2--5~Myrs exhibit high N/O  (due to N-rich WR stars) decreasing with time, as the ionised gas might diffuse into the cold neutral gas on longer timescales of $\sim$10--15~Myrs. 

Figure~\ref{Fig:enrich} shows 20 blue compact dwarf galaxies (12 from \citealt{Izotov06}; 1 each from \citealt{Brinchmann08},  \citealt{Guseva09} and \citealt{Izotov17}; and 5 from \citealt{Arellano-Cordova25}\footnote{Recently,  \citet{Arellano-Cordova25b} reported 10 galaxies with log(N/O)\,$>\!-1.1$. Among these, J0036-3333 is a well-known galaxy Haro 11, its resolved star-forming regions have been studied by \citet{James13b}, and none of its three star-forming knots have log(N/O)\,$>\!-1.1$. J0127-0619 and J0823+2806 are already included in this work as Mrk 996 and the SFG from \citet{Brinchmann08} respectively. J0808+3948 and J1253-0312, latter in common with \citet{Izotov06}, are AGN. The remaining 5 galaxies are plotted in Figure~\ref{Fig:enrich}.}) that have been found to have log(N/O)\,$>\!-1.1$ (red circles). Only three of these SFGs (J0519+0007, \citealt{Guseva09}; J1205+4551, \citealt{Izotov17}; J0944+3442, \citealt{Arellano-Cordova25b}) have 12+log(O/H)$<$8. Their log(N/O) values are among the lowest compared to the N-emitters in our sample. Note that, including Mrk 996 and these three SFGs, only four SFGs were previously known to be metal-poor (12+log(O/H)$<$8) N-emitters at low-z; our work adds a further 17 such metal-poor N-emitters in the local universe (see Table~\ref{tab:sample}).

We try not to include AGN at all in Figure~\ref{Fig:enrich}. There are 5 more blue compact dwarf galaxies that had previously been reported as SFGs with log(N/O)\,$>\!-1.1$ in the literature (3 from \citealt{Izotov06}; J0808+3948 from \citealt{Arellano-Cordova25b}; and HS~0837+4717 from \citealt{Pustilnik04}). However later, these are  confirmed to be AGN either from X-ray follow-up \citep[][]{Birchall20}, mid-IR variability \citep[][]{Aravindan24}, optical-continuum variability \citep{Burke21}, WISE colours \citep[][]{Lambrides19}, or spatially-resolved spectroscopy \citep[][]{Mezcua24}. In addition, we find one more  object from \citet{Izotov06} with log(N/O)\,$>\!-1.1$ that is part of DESI DR1 but noted as an AGN candidate due to its WISE colours.


\subsection{Comparison with high-redshift SFGs with log(N/O) determinations} 
\label{sec:high-z}

Figure~\ref{Fig:enrich} shows each of the high-z SFGs with log(N/O) determinations from optical N lines reported till date (blue squares). These include 5 SFGs from \citet{Stiavelli25} at z$\sim$3.2--6.3, 5 from \citet{Sanders24} at z$\sim$3.3--6, one from \citet{Marques-Chaves25} at z=6.2, one from \citet{Zhang25} at z=4.694, one from \citet{Rogers24} at z=2.963, 2 from \citet{Arellano-Cordova25} at z$\sim$5.2, 10 from \citet{Scholte25} at z$\sim$1.8--4.9, one from \citet{Welch24} at z=1.329, two star-forming regions in  an SFG at z=6.85 \citep{Scholtz25}, and the Lyman-continuum leaking star-forming region in the Sunburst Arc at z=2.37 \citep[][]{Welch25}. 11 of these determinations have log(N/O)$>-1.1$ (one from \citealt{Stiavelli25}; all five from \citet{Sanders24}; the one from \citealt{Zhang25}; the one from \citealt{Welch24}; the two from \citealt{Arellano-Cordova25}; and the star-forming region in the Sunburst Arc). J0217–0208 from \citet{Marques-Chaves25} at z=6.2 has the highest log(N/O)=$-0.3\pm0.1$ among these SFGs. Note that \citet{Cataldi25} also showed some additional N-emitters at cosmic noon with deep JWST/NIRSpec spectra.

The low-z N-emitters identified in this work have similar log(N/O) values with those at high-z, albeit at lower O abundances. Particularly the low-z N-emitters with high stellar-mass ($M_*>10^9$~M$_{\odot}$) also occupy similar mass range as the high-z SFGs. It thus seems reasonable to consider that the mechanisms driving such high log(N/O) values are similar among both samples (see Section \ref{sec:models} for discussion). 

Figure~\ref{Fig:enrich} also shows each of the high-z SFGs with log(N/O) determinations from UV N lines reported till date (orange diamonds and triangles). There are 7 such SFGs: A1703-zd6 \citep[z=7.0435,][]{Topping25}, RSCJ2248-JD \citep[z=6.1057,][]{Topping24}, GLASS150008 \citep[z=6.23,][]{Isobe23}, CEERS1019 \citep[z=8.68,][]{Marques-Chaves24}, GNz9p4 \citep[z=9.38,][]{Schaerer24}, GNz8LAE \citep[z=8.28,][]{Navarro-Carrera24}, and GNz9-0 \citep[z=9.43,][]{Curti24}. Also based on UV N lines, GNz11 at z=10.6 has been estimated to have log(N/O)$\geq-0.25$  \citep{Cameron23}, while GNz2 at z=12.34 has been estimated to have log(N/O)$\leq-0.4$ \citep{Castellano24}. 

It is notable from Figure~\ref{Fig:enrich} that all UV-based determinations of N/O result in high values are well above log(N/O)=$-1.1$. They are placed among the most extreme N-emitters identified in this work (No. 6 in Table ~\ref{tab:attributes} and Figure~\ref{Fig:stamps}). Only the nuclear region of Mrk~996 resembles CEERS~1019 in that it has log(N/O)=$-0.18\pm0.11$. Note that, however, it has been widely suggested \citep[e.g.][]{Stiavelli25} that the log(N/O) obtained from UV N lines are overestimated, potentially by $\sim$0.3--0.4 dex relative to the optical benchmark \citep{Martinez25}.  


\section{Probing the nature of extreme N-emitters at low redshift with GCE models} 
\label{sec:models}

The low-z N-emitters identified in this work span a range of masses and sSFR (Section~\ref{sec:mass}), but the most massive ones tend to have low sSFR and markedly low 12+log(O/H) as shown in Figure~\ref{Fig:mass_sfr}. The lower mass ones span a wider range of sSFR with low 12+log(O/H). We here construct GCE models to explain the log(N/O) vs 12+log(O/H) parameter space spanned by N-emitters at low-z identified in this work, which will help in understanding N-emitters at high-z.

Together with the latest nucleosynthesis yields, classical one-zone GCE models have been instrumental in developing our understanding of chemical enrichment processes in galaxies \citep[e.g.][]{kob11agb,Nomoto13,Matteucci21}. Such GCE models have primarily been constrained by  elemental abundance determinations of stars in the Milky Way \citep[e.g.][]{hayden15} showing that an interplay of core-collapse supernovae (CCSNe, including hypernovae), Type Ia supernovae (SNe Ia) and asymptotic giant branch (AGB) stars can explain the determined abundances of nearly all elements in the solar neighbourhood \citep{kob20sr}. The same chemical enrichment mechanisms but paired with a distinct star formation history \citep{Arnaboldi22,Kobayashi23} can reproduce the observed O \& Ar elemental abundances for planetary nebulae in the disc of the Andromeda galaxy \citep[][]{Bhattacharya22}. GCE models have also been constrained by elemental abundance determinations of stars in Milky Way dwarf spheroidal satellite galaxies  uncovering the potential role played by sub-Chandrasekhar mass SNe Ia (sub-Ch SNe Ia) in the chemical enrichment of such low-mass metal-poor galaxies \citep{kob20ia}. 

\citet{Kobayashi24} utilised GCE models, with an intermittent star formation history having two strong starbursts separated by a quiescent phase lasting $\sim$100~Myr and having outflows, to provide a possible explanation for the extreme N-enhancement of GNz11. It is important to note that the quiescent phase is not particularly required and the essential factor is the dual burst (see four other models in their Figure 4 of Appendix). Akin to what has been observed for NGC 5253 by \citet{Abril-Melgarejo24}, WR stars temporarily ($<1$~Myr) enhance N abundances in GNz11 following the second burst to become the dominant enrichment source. 

\begin{figure*}
\centering
\includegraphics[width=\textwidth]{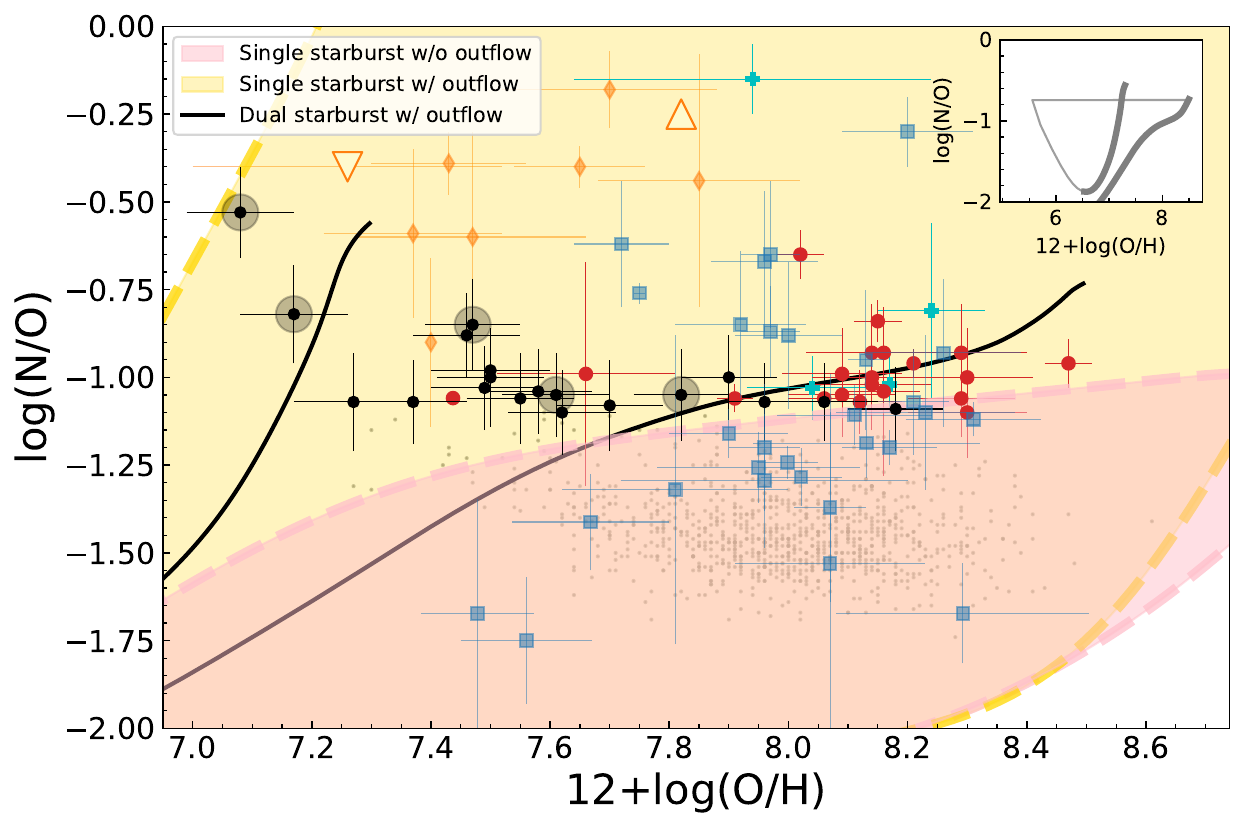}
\caption{Same as Figure~\ref{Fig:enrich} but with the locus traced by our GCE models marked (see Section~\ref{sec:models} for details). The 5 most massive extreme N-emitters (Table~\ref{tab:attributes}) are encircled. The inset shows 12+log(O/H) vs log(N/O) with an extended x-axis compared to the main plot, showing the complete locus traced by the dual starburst model. The locus positions accompanied by appreciable star-formation are marked by the thicker solid lines (see Section~\ref{sec:models} for details). 
} 
\label{Fig:model}
\end{figure*}

\subsection{Constructing GCE models} 
\label{sec:gce}

We construct a new set of GCE models using the GCE code by \citet{kob00} but including the latest nucleosynthesis yields of AGB stars, super-AGB stars, CCSNe \citep{kob20sr}, SNe Ia and sub-Ch SNe Ia \citep{kob20ia}, as well as WR stars \citep{Kobayashi24}.
The standard IMF from \citet{kro08} is adopted for $0.01M_\odot$ to $120 M_\odot$.
Note that, the SN Ia model is taken from \citet{kob09}, which includes a metallicity effect of Ch-mass explosion in single degenerate systems \citep{kob98}. Sub-Ch mass explosions from single and double degenerate systems are also included\footnote{Equation 2 of \citet{kob09} is used also for sub-Ch-mass SNe Ia but with different secondary mass ranges; 0.835--1.9$M_\odot$for single degenerate systems \citep{kob15} and 1.8--7.95$M_\odot$ for double degenerate systems, depending on the metallicity.}. However, they do not affect N/O ratios at all. It is important to note the model can well reproduce the observed elemental abundances (from C to Zn) in the solar neighbourhood \citep{kob20sr}.

The GCE models include exponential inflow (with timescale $\tau_{\rm i}$), SFR proportional to the gas fraction ($\tau_{\rm s}$), and outflow rate also proportional to the gas fraction ($\tau_{\rm o}$). See \citet{Kobayashi25} for the formula (Eqs.9,10,12). 
Three set of models are constructed including a multiple starburst. Massive stars formed in the first burst in each of the models explode as CCSNe enriching the ISM with O, while N is contributed at early times from WR stars and at later times from AGB stars. This causes the increase of N/O toward higher metallicities (e.g. O abundances) in all models. The model predictions are shown in Figure~\ref{Fig:model} and can be summarised as follows:

\begin{itemize}
    \item {\it Single starburst without outflow}: The pink shaded area in Figure~\ref{Fig:model} shows the range of single-burst models for 3 Gyr after the onset of star formation, with varying star formation timescale from $\tau_{\rm s}\!=$\,0.1 to 30 Gyr. The model with $\tau_{\rm s}\!=$\,0.1 Gyr corresponds to the highest star formation efficiency among the models and is located at the metal-rich edge.  The inflow timescale is fixed to $\tau_{\rm i}=0.1$ Gyr as it does not change the elemental abundance tracks at all. 

    \item {\it Single starburst with outflow}: The yellow shaded area in Figure~\ref{Fig:model} shows single-burst strong-outflow models with $\tau_{\rm o}=\tau_{\rm i}=0.1$ Gyr, again varying $\tau_{\rm s}$ from 0.1 to 30 Gyr, which can produce high N/O ratios over a wide range of O/H. With outflows, N/O ratios becomes higher due to loss of O and can explain {\it all} high N/O values in SFGs.  Note that, with a longer $\tau_{\rm o}$ value, i.e. weaker outflow, the yellow area approaches to the pink area. 
    With these outflows, N/O becomes higher mainly with the delayed N enrichment from AGB stars. 
    
    \item {\it Dual starburst with outflow}: There is another way to produce high N/O at low metallicities -- a dual starburst \citep{Kobayashi24}. The black solid lines in Figure~\ref{Fig:model} show the star formation (observable) episodes of a dual starburst model;
    both bursts have the same star-formation timescale $\rm\tau_s=10$~Gyr
    but the infall timescales are $\rm\tau_i=0.1$ and 0.2 Gyr, respectively, and the outflow timescales are $\rm\tau_o=0.5$ and 0.1 Gyr, respectively, which results in
    the second burst being only a third of the SFR of the first one. 
    The full evolutionary track is shown in the inset of Figure~\ref{Fig:model} in the 12+log(O/H) vs log(N/O) plane, where 
    the thick solid lines show the locus associated with each of the bursts of star formation that have observable number of stars. The loop is the signature of ISM dilution following gas infall prior to the start of the second burst. During the first burst, even with moderate outflow, the model locus traces an increasing N/O with rapidly increasing O/H (the right-most thick line). 
    Then, the secondary inflow of primordial gas dilute the ISM, suddenly decreasing O/H (the narrow horizontal line). CCSNe from the secondary star burst increase O, causing the rapid decrease of N/O while increasing O/H (the left-most narrow line). Finally AGB stars from secondary star burst enhance N, causing the increase of N/O (the inner thick line). 
\end{itemize}

Note that the dual burst model track is similar but not as bursty as for the one for GN-z11 in \citet{Kobayashi24}; the extra enhancement from WR stars before the O production from CCSNe is included but is not seen. Also for $\sim$40 Myrs after the onset of the first burst, AGB stars do not contribute in chemical evolution, which is important to note for very high-z galaxies.

\subsection{GCE model-based inference} 
\label{sec:infer}

The aforementioned models provide considerable insight into the nature of extreme N-emitters, especially at low-z. 

The single starburst models without outflow (Figure~\ref{Fig:model}; pink) can well reproduce the local scaling relation (see Figure~\ref{Fig:enrich}) where observed N/O ratios increase with increasing O abundances, as well as the bulk of SFGs with O \& N abundances determined in this work. However, they cannot explain the N-emitters, especially at low metallicities, 12+log(O/H)$<8$.

Among our N-emitters, those with relatively low stellar mass (approx. M$_*<10^9$~M$_{\odot}$; see Table~\ref{tab:attributes}) can be explained with the single starburst models with outflow (Figure~\ref{Fig:model}; yellow). As they are dwarf galaxies, strong outflow from star-formation activity is expected due to their small gravitational potential well. 
It is important to note that we do not necessarily expect 
ongoing outflow signature  present in the spectra of these SFGs. 
Our GCE models show that outflows during the evolution of these galaxies result in high N/O at a given O/H with delayed N-enrichment from AGB stars.
However, for the non-AGN high-mass galaxies such a strong outflow is not expected.  

For the massive N-emitters in our sample (approx. M$_*>10^9$~M$_{\odot}$; see Table~\ref{tab:attributes}), with the five most massive ones marked in Figure~\ref{Fig:model}, dual starburst with outflow is more plausible. This particular model assumes the peak SFR lower in the second burst, which is also consistent with cosmological growth of galaxies.
Generally, dual starburst with similar outflow 
can explain the high N/O ratios too. Having already experienced a previous burst of star-formation, the galaxy formed in the dual starburst case will likely be more massive and thus have lower sSFR, consistent with the properties of extreme N-emitting SFGs in our sample (see Section~\ref{sec:mass} and Figure~\ref{Fig:frac}c). 

Given the aforementioned scenarios for low-z galaxies, 
we find that sustained N-enhancement by AGB stars, in conjunction with presence of outflows during the evolution of the galaxy, can explain the observed high N/O ratios of low-z N-emitters. Such a scenario may be applicable to the relatively massive high-z N-emitters at $z<9$, 
when there has been sufficient time ($>$0.5~Gyr) for N-enrichment by AGBs, regardless of whether single, dual or multiple episodes of starburst have occurred. The high-z N-emitters at $z>9$, 
with UV-based N-abundances determined, require more rapid enrichment modes ($<$0.5~Gyr) like WR stars.

As we move to higher redshifts, galaxies have had shorter lifespans and so signatures of ongoing outflows may be present in the highest-z N-emitters (albeit at z$<9$). Indeed J0217-0208 from \citet{Marques-Chaves25} at z=6.2 having log(N/O)$=-0.3 \pm 0.1$, the highest of any high-z N-emitter with optical N-lines, shows signatures of strong outflows (broad components in several rest-optical emission lines combined with ALMA detections of a massive, extended cold dust reservoir), consistent with N-enhancement from AGB stars in systems with strong outflows as per our proposed GCE models. The same is true for CEERS1019 at z=8.68 where outflows have recently been identified with NIRSPEC IFS \citep{Zamora25}. 

Note again that WR-stars are also included but their instantaneous N-enhancement is not visible in the GCE models of this paper
\footnote{As a diagnostic test, we re-compute the GCE model predictions by setting WR contribution to zero, and find no discernible change in the predicted N/O vs O/H plots for any of our models.}. 
There is no detection of the HeII$\lambda$ 4686 \AA~line, which is expected to accompany the characteristic blue bump in galaxies with WR stars \citep[e.g.][]{Lopez-Sanchez09,Curti25}, in any of our extreme N--emitters (see Figures~\ref{Fig:exp_spec}~\&~\ref{Fig:app_spec_1}~\&~\ref{Fig:app_spec_2}). The HeII$\lambda$ 4686 \AA~line also remains undetected in the stacked spectrum of our extreme N-emitters (see Appendix~\ref{appendix: stack}). However, given the faintness of this line even for known galaxies with WR stars \citep[e.g.][]{Lopez-Sanchez09}, any contribution of WR-stars to the N-enhancement for the low-z SFGs remains unclear. 



\section{Conclusion}
\label{sec:conc}

We obtain direct N \& O abundances for a sample of 944 SFGs at $z<$0.5 in DESI DR1, and find that 19 of these SFGs are N-emitters with log N/O\,$>-1.1$ (Section~\ref{sec:enrich}).
This is a five-fold increase in the number of metal-poor (12+log(O/H)$<8$) low-z N-emitters (17 from our sample), as only four such SFGs were known previously (Section~\ref{sec:low-z}). Four of the N-emitters identified in this work have the lowest 12+log(O/H) values determined till date for such galaxies (Figure~\ref{Fig:enrich}). The most extreme N-emitter in our sample (No. 6 in Table~\ref{tab:attributes} with log(N/O)=$-0.53\pm0.13$) also has the lowest 12+log(O/H) with $7.08\pm0.09$ and the highest stellar mass, log(M$_{*}$/M$_{\odot}$)=9.95 ± 0.13. 

The homogeneous sample allows us to compute the N-emitter fraction of $2.21\pm0.91$\%. 
These N-emitters span a wide range of SFR values, but a marked number of them have higher stellar masses and low sSFR, as well as low O abundances (Figure~\ref{Fig:mass_sfr}). In fact, the fraction of N-emitters increases with increasing mass, decreasing sSFR and decreasing O abundance (Figure~\ref{Fig:frac}). 

Constructing new set of GCE models (Section~\ref{sec:models}), we show that single starburst models without outflows cannot explain the high log(N/O) values of the low-z N-emitters at 12+log(O/H)\,$<\!8$, and strong outflows are required (Figure~\ref{Fig:model}). However, such strong outflows are not very likely in our relatively massive non-AGN SFGs 
(approx. M$_*\!>\!10^9$~M$_{\odot}$; see Table~\ref{tab:attributes}).   For these galaxies, dual starbursts with outflows better explains the high log(N/O) values. 

We thus show that extreme N-emitters at low O abundances, such as those observed at high-z, are also present in the local universe, allowing us to gain significant insight on their chemical enrichment mechanisms. N-enrichment from AGB stars, coupled with strong outflows sometime during the lifetime of a galaxy in single and dual starbursts, are the likely drivers of high log(N/O) in low-z N-emitters. Such enrichment mechanisms may also explain the high log(N/O) in high-z SFGs out to z$\sim$9.
A larger homogeneous sample
across cosmic time with multi-object spectrographs on 8--10m class ground-based telescopes, such as Subaru’s Prime Focus Spectrograph and VLT's Multi Object Optical and Near-infrared Spectrograph, are required to probe the origin of N-emitters, as well as targeted observations with the James Webb Space Telescope.


\section*{Acknowledgements}

We thank the anonymous referee for their valuable suggestions. CK acknowledges funding from the UK Science and Technology Facility Council through grant ST/Y001443/1. This research used data obtained with the Dark Energy Spectroscopic Instrument (DESI). DESI construction and operations is managed by the Lawrence Berkeley National Laboratory. This material is based upon work supported by the U.S. Department of Energy, Office of Science, Office of High-Energy Physics, under Contract No. DE–AC02–05CH11231, and by the National Energy Research Scientific Computing Center, a DOE Office of Science User Facility under the same contract. Additional support for DESI was provided by the U.S. National Science Foundation (NSF), Division of Astronomical Sciences under Contract No. AST-0950945 to the NSF’s National Optical-Infrared Astronomy Research Laboratory; the Science and Technology Facilities Council of the United Kingdom; the Gordon and Betty Moore Foundation; the Heising-Simons Foundation; the French Alternative Energies and Atomic Energy Commission (CEA); the National Council of Humanities, Science and Technology of Mexico (CONAHCYT); the Ministry of Science and Innovation of Spain (MICINN), and by the DESI Member Institutions: www.desi.lbl.gov/collaborating-institutions. The DESI collaboration is honored to be permitted to conduct scientific research on I’oligam Du’ag (Kitt Peak), a mountain with particular significance to the Tohono O’odham Nation. Any opinions, findings, and conclusions or recommendations expressed in this material are those of the author(s) and do not necessarily reflect the views of the U.S. National Science Foundation, the U.S. Department of Energy, or any of the listed funding agencies.

\section*{Data Availability}

Based on tabulated data publicly available at the \href{https://data.desi.lbl.gov/doc/releases/dr1}{DESI DR1 Data Release}. GCE models can be shared upon reasonable request. 



\bibliographystyle{mnras}
\bibliography{oar} 

\appendix

\section{Spectra of extreme N-emitters}
\label{appendix: spectra}

\begin{figure*}
\centering
\includegraphics[width=0.9\textwidth]{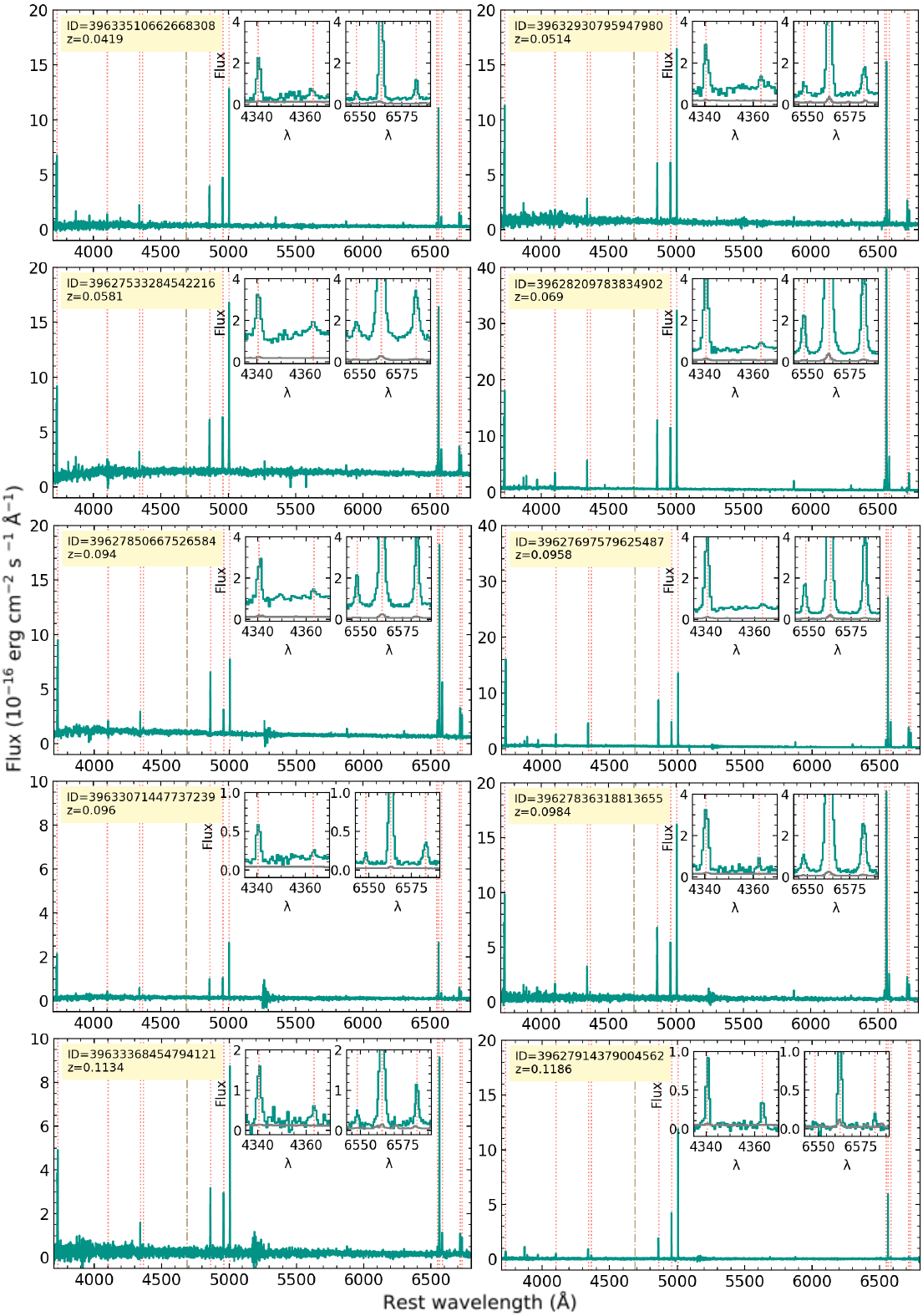}
\caption{Same as Figure~\ref{Fig:exp_spec} but showing the DESI spectra for the N-emitters No. 2--11 in Table~\ref{tab:attributes}. 
The insets clearly show the detections of the [OIII]$\lambda$ 4363 \AA~and [NII]$\lambda$ 6583 \AA~lines in all sources, with the [NII]$\lambda$ 6548 \AA~line also detected in all but one source. Note also that the axis scales vary for each spectra (also for the insets) to accommodate the fluxes of the brightest lines.} 
\label{Fig:app_spec_1}
\end{figure*}

\begin{figure*}
\centering
\includegraphics[width=0.91\textwidth]{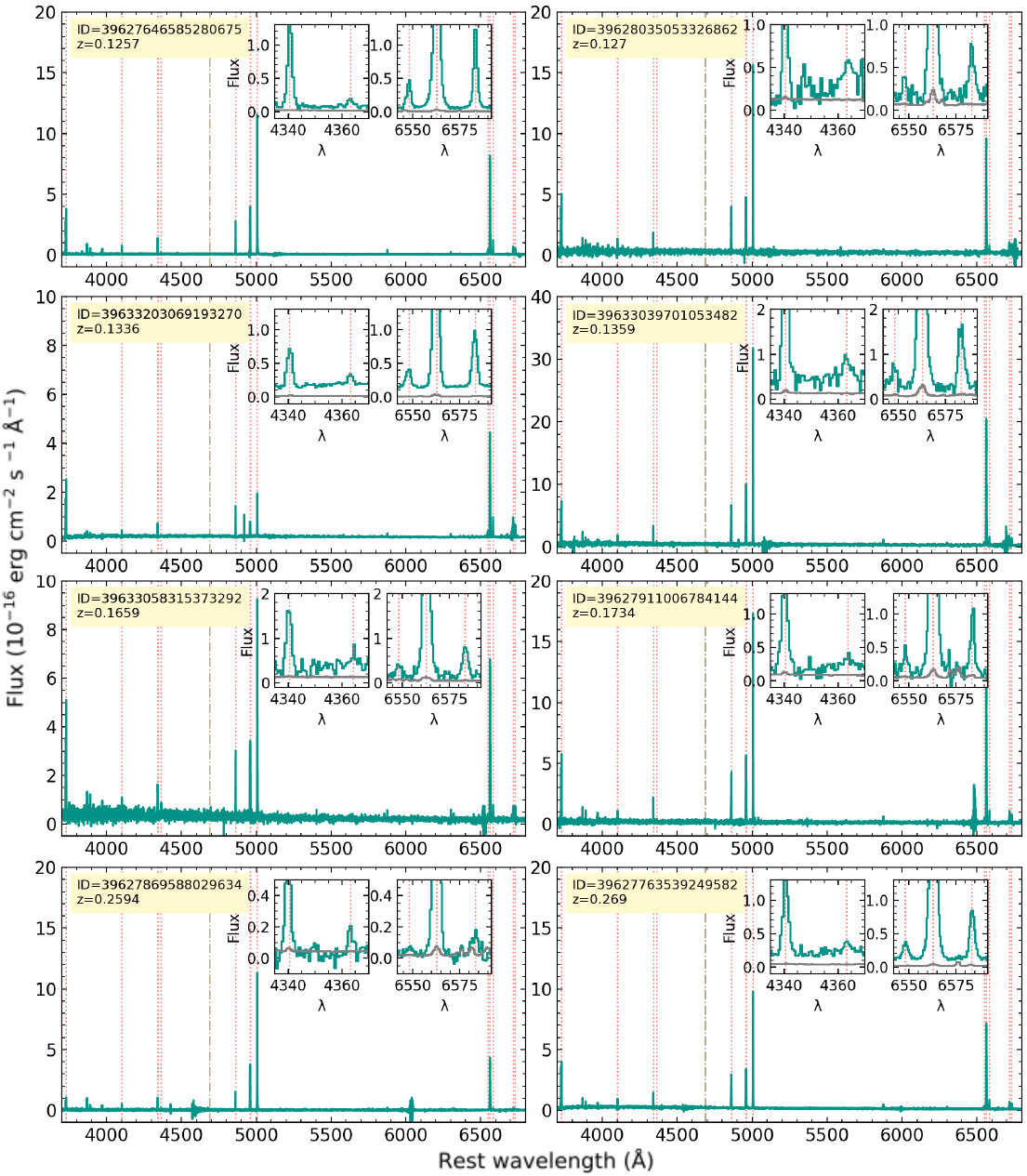}
\caption{Same as Figure~\ref{Fig:app_spec_1} but showing the DESI spectra for the N-emitters No. 12--19 in Table~\ref{tab:attributes}. The insets clearly show the detections of the [OIII]$\lambda$ 4363 \AA~and [NII]$\lambda$ 6548, 6583 \AA~lines in all sources.} 
\label{Fig:app_spec_2}
\end{figure*}

As in Figure \ref{Fig:exp_spec}, the DESI spectra for the rest of N-emitters (No. 2--18 in Table~\ref{tab:attributes}) are shown in Figures~\ref{Fig:app_spec_1}~\&~\ref{Fig:app_spec_2}.  The [NII]$\lambda$ 6548, 6583 \AA~lines are detected for all but one source (No. 11 at z=0.1186 in Figure~\ref{Fig:app_spec_1}) where only the [NII]$\lambda$ 6583 \AA~line is detected. The fluxes of these lines, tabulated in the VAC, are thus reliably detected for each source, where present, in the DESI spectra, using the automated procedure outlined in \citet{Zou24}. This is further shown by the consistency in our determined abundances compared to \citet{Zinchenko24} in Section~\ref{sec:lit}. We reiterate that the spectra are shown only for visual confirmation, and the required tabulated line fluxes from the VAC are utilised for abundance determination in this work.

\section{Stacked spectra of extreme N-emitters}
\label{appendix: stack}

\begin{figure}
\centering
\includegraphics[width=\columnwidth]{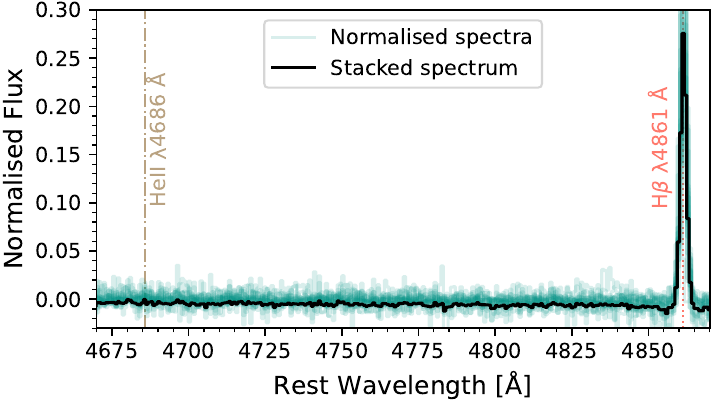}
\caption{The stacked spectrum is shown in black while the normalised spectra of the 19 N-emitters in Table~\ref{tab:attributes} are shown in green (they overlap with each other in the noise but the H-$\rm\beta$ line is clear).   The wavelength range covers the marked position of the undetected HeII$\lambda$ 4686 \AA~line and the clear H-$\rm\beta$ line.} 
\label{Fig:stack}
\end{figure}

While the HeII$\lambda$ 4686 \AA~line remains undetected for any of our N--emitters (see Figures~\ref{Fig:exp_spec}~\&~\ref{Fig:app_spec_1}~\&~\ref{Fig:app_spec_2}), it does not rule out the presence of WR stars in these SFGs as this line is intrinsically faint even in confirmed WR galaxies (the flux of this line can be $\sim$0.0025 times that of the [OIII]$\lambda$ 5007 \AA~line; \citealt{Lopez-Sanchez09}). We further computed the stacked spectrum from the individual DESI spectra of the 19 N-emitters, adopting the spectral resampling using SpectRes (\citealt{Carnall17}) and normalised to their [OIII]$\lambda$ 5007 \AA~line fluxes. Figure~\ref{Fig:stack} shows the stacked spectrum as well as the normalised spectra of the 19 N-emitters covering the wavelength range of the HeII$\lambda$ 4686 \AA~and H-$\rm\beta$ lines. It is clear that HeII$\lambda$ 4686 \AA~line is not detected in the stacked spectrum either, even though the noise in the spectrum is clearly reduced compared to the individual normalised spectra. The 3$\sigma$ detection limit at the position of the HeII$\lambda$ 4686 \AA~line (relative to the [OIII]$\lambda$ 5007 \AA~line) is 0.018, so indeed even known WR galaxies (e.g. with this ratio being $\sim$0.0025; \citealt{Lopez-Sanchez09}) would remain undetected. Thus we can neither confirm nor rule out the presence of WR stars even with the stacked spectra.

\bsp	
\label{lastpage}
\end{document}